# AcerDET-2.0[1]: a particle level fast simulation and reconstruction package for phenomenological studies on high $p_T$ physics at LHC.

**Patryk Mikos**
*Theoretical Computer Science, Jagellonian University;*
*30-348 Krakow, ul. Lojasiewicza 6, Poland.*
**Elżbieta Richter-Wąs**
*Institute of Physics, Jagellonian University;*
*30-348 Krakow, ul. Lojasiewicza 11, Poland.*

## Abstract

The fortran version of the `AcerDET` package has been published in [1], and used in the multiple publications on the predictions for physics at LHC. The package provides, starting from list of particles in the event, the list of reconstructed jets, isolated electrons, muons, photons and reconstructed missing transverse energy. The `AcerDET` represents a simplified version of the package called `ATLFAST`, used since several years within ATLAS Collaboration. In the fast simulation implemented in `AcerDET`, some functionalities of `ATLFAST` are absent, but the most crucial detector effects are implemented and the parametrisations are largely simplified. Therefore it is not representing details neither of ATLAS nor CMS detectors.

This short paper documents a new C++ implementation of the same algorithms as used in [1]. We believe that the package can be well adequate for some feasibility studies of the high $p_T$ physics at LHC and at planned ppFCC. The further evolution of this code is planned.

---

[1]Package is available for the web page `http://erichter.home.cern.ch/erichter/AcerDET.html`

# PROGRAM SUMMARY

*Title of the program:* **AcerDET version 2.0**
*Operating system:* Linux
*Programming language:* C++ .
*External libraries:* HepMC, ROOT.
*Size of the compressed distribution directory:* about 25 kB.
*Nature of physical problem:* A particle level fast simulation and reconstruction package. The package provides, starting from the list of particles in the event, list of reconstructed jets, isolated electrons, muons and photons and reconstructed missing transverse energy. The algorithmic part is just rewritten from fortran version of the package `AcerDET-1.0` [1]. The default input is generated event in format of HepMC structure [2], output is in form of ROOT file [4], containing tree with reconstructed objects and set of control histograms. Distribution version includes also example of the main program for execution with `PYTHIA 8.2` generator [3]. Implemented set of parametrisations is not representing in details ATLAS or CMS detectors, some of them are simple and can be considered rather as place-holders for future adaptation of any detector. Nevertheless, we believe that the package will be well adequate for some feasibility studies on the high $p_T$ physics at LHC.

# 1 Introduction

The potential of the LHC detectors for physics at high $p_T$ is very rich, as first three years of data taking during so called `Run I` have shown.

The discovery of the Higgs boson, searches results for Supersymmetry particles, Exotic particles, New Vector Bosons are very impressive. All these is thanks to the high sensitivity of the detectors in terms of acceptance and identifying efficiencies to variety of signatures: photons, electrons, muons, multi-b-jets, $\tau$-jets, missing transverse energy.

Those high sensitivities allow to study very exclusive signatures; the discovery potential is limited in some cases by the rare background processes. It is evident, that the multi-jet and multi-b-jet production in association with known vector bosons, W and/or Z, or with top-quark pair is the the serious background to the several observables as well. The importance of effect beyond the leading order: effects coming from the finite width or angular spin correlations (eg. from the intermediate resonances decays) is becoming more evident. Limitations and complementarity of the matrix element and parton shower predictions for the diversity of signatures need to be investigated.

The package for particle-level simulation and reconstruction is one of the intermediate steps between simple parton-level analysis and very sophisticated and CPU consuming full detector simulation. The package provides, starting from list of particles in the event, list of reconstructed jets, isolated electrons, muons and photons and reconstructed missing transverse energy. It can serve for phenomenological feasibility studies on the prospects for observability of a given signature. Such package can be useful for the dedicated comparisons between matrix element and parton shower predictions. In that case, one can compare experimental signatures (reconstructed jets, leptons, photons) to quantify size of the discrepancies between different predictions in a straightforward way.

The `Fortran` version of presented here package [1] has played such role since several years, more than 80 papers published in the last 10 years was using it or this purpose [2]. Here, we present the new `C++` implementation, where all the algorithmic components have been preserved, changed is format of the input event. Now comonly accepted `HepMC` and the output format of `ROOT` tree with reconstructed objets and control histograms is introduced.

The package simulates some key features of the LHC detectors like ATLAS and CMS. It is based on the calorimetric energy deposition for jets reconstruction and tracks reconstruction for electrons and muons. It takes into account very high granularity of the electromagnetic calorimetry for the photon reconstruction. The missing transverse energy is calculated for the total energy balance of the reconstructed objects. The capability for the identification of b-jets and $\tau$-jets is also explored. Implemented set of parametrisations is not representing in details performance of neither ATLAS nor CMS detectors. Nevertheless, we believe that the package is still adequate for feasibility studies on the high $p_T$ physics at LHC and offer starting option also for future FCC program.

The paper is organised as follows. In Section 2 we discuss (recall) algorithms used for objects reconstruction and show benchmarking distributions. Given the Higgs boson discovery at mass 125 GeV, benchmark which were published in [1] are now updated to this value of teh Higgs mass. Also `Pythia 8.1` showering model used here are different than what was used for [1], hence some benchmark figures are different despite same algorithms used for fast simulation and reconstruction. Section 3 gives an outlook. We don't recall comparison with ATLAS performance numbers (Appendix A of [1]) as those are not updated as well with the better inderstanding of the detector using Run I data and are much more now detailed and analysis dependent. One should just to look for the publications related to the signature of interest. It was however checked that numbers from Appendix A of [1], are reproduced by AcerDET-2.0 code.

In the Appendixes A-F, we give more technical details concerning input parameters and output structure. In Appendix G we show control output logfile.

## 2 Simulation and reconstruction

Fully or partially generated event i.e. event generated including or not QED/QCD initial and final state radiation, fragmentation, hadronisation and decays of unstable particles can by analysed by this package. The list of the partons/particles of the generated event should be stored in the `HepMC` format. It is then copied into internally used vector of particles, with added few flags i.e. additional information regarding particles status and its type. This information is used by the `AcerDET` algorithms for event-be-event simulation and reconstruction.
Events simulation is limited to the following steps:

- Deposition of the particles energies in the calorimetric cells.

- Smearing of the energy of electrons, photons, muons with the parametrised resolutions.

- Smearing of the energy of hadronic clusters and not-clustered cells with the parametrised resolution.

Events reconstruction is limited to the following steps:

- Reconstruction of the calorimetric clusters.

- Verification of the isolation criteria for electrons/photons/muons.

- Rejection of clusters associated with electrons and photons.

- Acceptation of remaining clusters as hadronic jets and identification (labeling) of those associated with b-quarks, c-quarks, tau-leptons.

- Jets energy calibration.

There is no clear separation between the *simulation* and *reconstruction* parts in the structure of algorithms. A larger blocks (`algorithms`) may realise both tasks.

As a final result provided are four-momenta of reconstructed objects: electrons, photons, muons, labeled and calibrated jets and calculated is total missing transverse energy in analysed event.

## 2.1 Calorimetric clusters

The transverse energy of all undecayed particles stored in the event record, (`InputRecor:Particles`, except of muons, neutrinos and other *invisible* particles[2] eg. the lightest SUSY particle, are summed up in the map of calorimetric cells with a given granularity in $(\eta \times \phi)$ coordinates (default: $0.1 \times 0.1$ for $|\eta| < 3.2$ and $0.2 \times 0.2$ for $|\eta| > 3.2$, with the calorimetric coverage up to $|\eta| = 5.0$). As an effect of the solenoidal magnetic field in the inner part of the detector we assume that the $\phi$ position of charged particles with transverse momenta above the threshold (default: $p_T > 0.5$ GeV) will be shifted as parametrised in function `Particle::foldPhi`. The contribution from all charged particles with transverse momenta below that threshold is neglected.

All calorimetric cells with the transverse energy greater than a given threshold (default: $E_T > 1.5$ GeV) are taken as possible initiators of clusters. They are scanned in order of decreasing $E_T$ to verify whether the total $E_T$ summed over all cells in a cone $\Delta R = \sqrt{\Delta^2\eta + \Delta^2\phi}$ exceeds the minimum required threshold for the reconstructed cluster (default $E_T > 5$ GeV). Cells with deposited transverse energy below the threshold (default: $E_T = 0$) are not accounted for. As a coordinates $(\eta^{clu} \times \phi^{clu})$ of the reconstructed cluster taken are the coordinates of the bary-center of the cone weighted by the cells $E_T$ for all cells inside the cone around the initiator cell. Information about energy deposition in cells is stored in `OutputRecord:Cells`

All reconstructed clusters are stored in the container of `ClusterData` objects, `OutputRecord:Clusters`. Fig. 1 shows the $\Delta\eta$ (left) and $\Delta\phi$ (right) distribution between the reconstructed bary-center of particles falling within the geometrical cluster cone and the reconstructed cluster position.

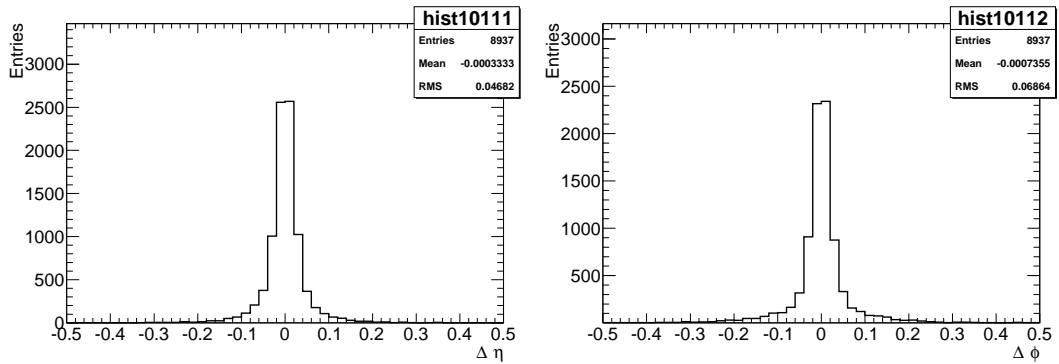

Figure 1: *The $\Delta\eta$ (left) and $\Delta\phi$ (right) distribution between the reconstructed bary-center of particles falling within the cluster cone and the reconstructed cluster position. Shown are results for generated $gg \to H, H \to u\bar{u}$ process with $m_H = 125~GeV$.*

---

[2]User may want to make any particle invisible to the detector. For this one should overwrite its *pdgId* code in the `vector<Particle>` to that specified for the invisible particles in `acerdet.dat` file.

## 2.2 Isolated muons

Algorithm reconstructing isolated muons uses information of the generated muons, reconstructed calorimetric clusters and the cells map.

Isolated muon candidates are searched for in the container `InputRecord.particles`. The inverse muon four-momentum is smeared according to the Gaussian resolution parametrised with function `Smearing::forMuon` (default: $\sigma = 0.05\% \cdot p_T$). The muon direction remains unsmeared.

For all muons which pass selection criteria in $p_T$ and $\eta$ (default: $p_T > 6$ GeV and $|\eta| < 2.5$ GeV), isolation criteria, in terms of the distance from calorimetric clusters and of maximum transverse energy deposition in cells in a cone around the muon, are then applied (defaults: separation by $\Delta R > 0.4$ from other clusters and $\sum E_T < 10$ GeV in a cone $\Delta R = 0.2$ around the muon). All muons passing the isolation criteria are stored in the container `OutputRecord.Muons` and those not passing are stored in `OutputRecord.NonisolatedMuons`.

As a control physics process the $gg \to H \to ZZ^* \to 4\mu$ production with the Higgs boson mass of 125 GeV is used. The default isolation selection has 97.8% efficiency for muons passing kinematical selection with $p_T^{\mu_1,\mu_2} > 20$ GeV and $p_T^{\mu_3,\mu_4} > 7$ GeV. As the predicted intrinsic width of the Higgs boson of this mass is very small, the resolution is completely dominated by the resolution assumed for the single muon reconstruction. Fig. 2 (left) shows the reconstructed distribution of the 4-muon system. The assumed single muon transverse momenta resolution of $\sigma = 0.05\% \cdot p_T$ leads to the $\sigma_m = 1.57$ GeV resolution for the invariant mass of the four-muon system originating from the $H \to ZZ^* \to 4\mu$ decay.

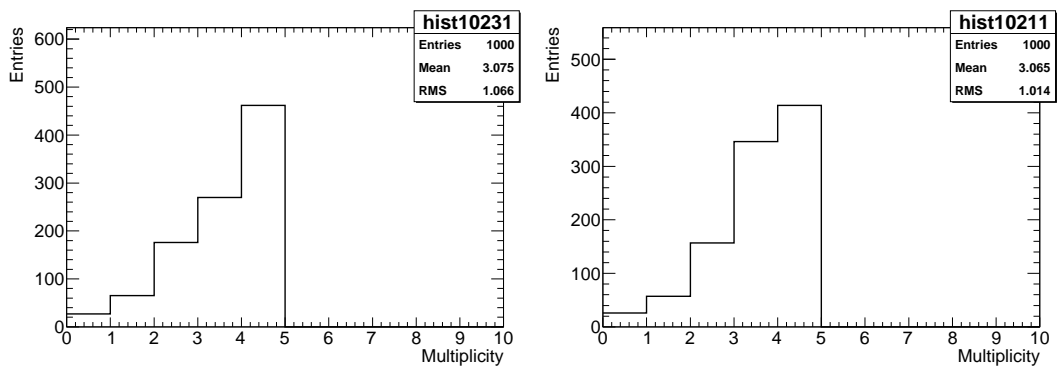

Figure 2: *Left: The multiplicity of hard-process isolated muons (left) and reconstructed isolated muons (right) for $gg \to H, H \to ZZ^* \to 4\mu$ events with $m_H = 125$ GeV.*

## 2.3 Isolated electrons

Algorithm reconstructing isolated electrons uses information of the generated electrons, reconstructed calorimetric clusters and the cells map.

Isolated electron candidates are searched for in the `InputRecord.particles`. The electron four-momentum is smeared according to the Gaussian resolution parametrised with function `Smearing::forElectron` (default: $\sigma = 12\%/\sqrt{E}$). The electron direction remains unsmeared.

For all electrons which pass selection criteria in $p_T$ and $\eta$ (default: $p_T > 5$ GeV and $|\eta| < 2.5$ GeV), the associated reconstructed calorimeter cluster is identified (default: $\Delta R_{e,cluster} < 0.1$). Electron isolation criteria, in terms of the distance from other clusters and of maximum transverse energy deposition in cells in a cone around the electron, are then applied (defaults: separation by $\Delta R > 0.4$ from other clusters and $\sum E_T < 10$ GeV in a cone $\Delta R = 0.2$ around the electron). All electrons passing the isolation criteria are stored in `OutputRecord.Electrons` and the clusters associated with them are removed from the `OutputRecord.Clusters`.

As a control physics process the $gg \to H \to ZZ^* \to 4e$ production with the Higgs boson mass of 125 GeV is used. The default isolation selection has 95.2% efficiency for electrons passing kinematical selection with $p_T^{e_1,e_2} > 20$ GeV and $p_T^{e_3,e_4} > 7$ GeV. As the predicted intrinsic width of the Higgs boson of this mass is very small, the resolution is completely dominated by the resolution assumed for the single electron reconstruction. Fig. 3 (left) shows the reconstructed distribution of the 4-electron system. The assumed single electron energy resolution of $\sigma = 12\%/\sqrt{E}$ leads to the $\sigma_m = 1.26$ GeV resolution for the reconstructed invariant mass of the four-electron system originating from the $H \to ZZ^* \to 4e$ decay.

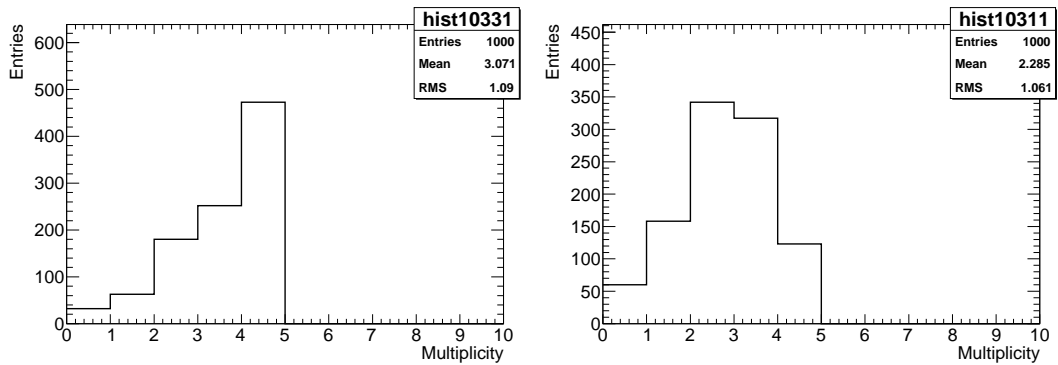

Figure 3: *Left: The multiplicity of hard-process isolated electrons (left) and reconstructed isolated electrons (right) for $gg \to H, H \to ZZ^* \to 4e$ events with $m_H = 125$ GeV.*

## 2.4 Isolated photons

Algorithm reconstructing isolated photons uses information of the generated photons, reconstructed calorimetric clusters and the cells map.

Isolated photon candidates are searched for in the `InputRecord.particles`. The photon four-momentum is smeared according to the Gaussian resolution parametrised with function `Smearing::forPhoton` (default: $\sigma = 10\%/\sqrt{E}$). The photon direction remains unsmeared.

For all photons which pass selection criteria in $p_T$ and $\eta$ (default: $p_T > 5$ GeV and $|\eta| < 2.5$ GeV), the associated reconstructed calorimeter cluster is identified (default:

$\Delta R_{\gamma,cluster} < 0.1$). Photon isolation criteria, in terms of the distance from other clusters and of maximum transverse energy deposition in cells in a cone around the photon, are then applied (defaults: separation by $\Delta R > 0.4$ from other clusters and $\sum E_T < 10$ GeV in a cone $\Delta R = 0.2$ around the photon). All photons passing the isolation criteria are stored in `OutputRecord.Photons` and the clusters associated with them are removed from the `OutputRecord.Clusters`.

As a control physics process the Standard Model $gg \to H \to \gamma\gamma$ production with the Higgs boson mass of 125 GeV is used. The isolation selection has 98.0% efficiency for photons passing the kinematical selection of $p_T^{\gamma_1} > 40$ GeV and $p_T^{\gamma_2} > 25$ GeV. As the predicted intrinsic width of the Higgs boson of this mass is much below 1 GeV, the resolution of the reconstructed invariant mass of the di-photon system is completely dominated by the resolution assumed for the single photon reconstruction. Fig. 4 (left) shows the reconstructed distribution of the di-photon system of photons passing selection criteria. The assumed energy resolution of $\sigma = 10\%/\sqrt{E}$ leads to the $\sigma_m = 0.85$ GeV resolution for the reconstructed invariant mass of the di-photon system originating from the $H \to \gamma\gamma$ decay.

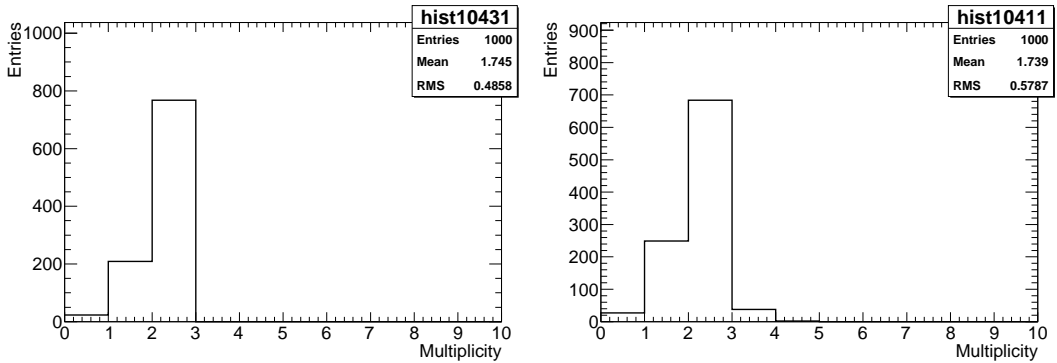

Figure 4: *Left: The multiplicity of hard-process isolated photons (left) and reconstructed isolated photons (right) for $gg \to H, H \to \gamma\gamma$ events with $m_H = 125$ GeV.*

## 2.5 Jets

Clusters which have not been selected as associated with electrons or photons are smeared with Gaussian resolution parametrised in function `Smearing::forHadron` (default: $\sigma = 50\%/\sqrt{E}$ and $100\%/\sqrt{E}$).

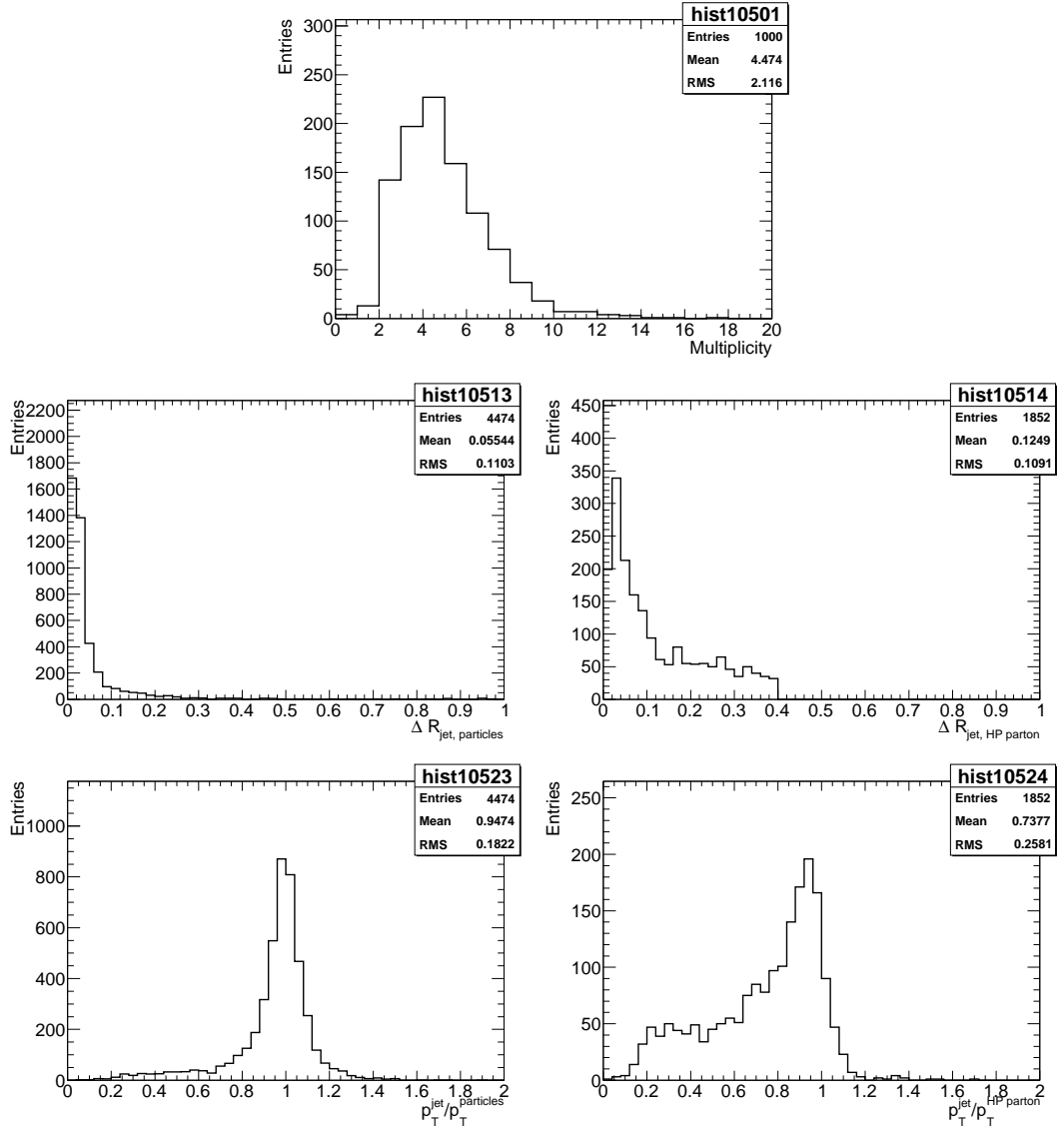

Figure 5: *Multiplicity of jets (top), $\Delta R$ cone distance between reconstructed jet and barycenter of particles (middle-left) and hard-process parton (middle right). The ratio of $p_T^{jet}/p_T^{particles}$ (bottom-left) and $p_T^{jet}/p_T^{HP parton}$ (bottom-right) for $gg \to H, H \to u\bar{u}$ and $m_H = 125~GeV$.*

If the non-isolated muon falls into the cone of a cluster its 4-momenta is added to the cluster 4-momenta and the cluster direction is recalculated. The resulting clusters are classified as jets if their transverse momentum is greater than a given threshold (default: $p_T > 15$ GeV). They are removed from the `OutputRecord.Clusters` and stored in the common `OutputRecord.Jets`.

The jets reconstruction efficiencies and di-jet mass resolution have been studied using control physics process of the $gg \to H$ production with $m_H = 125$ GeV and forcing the Higgs boson decay into specific partons, namely $H \to u\bar{u}$, $H \to c\bar{c}$ and $H \to b\bar{b}$. We used also process $gg \to H \to \tau\tau$ for estimating tau-jet reconstruction efficiency (we forced tau-leptons decay into hadrons). For estimating reconstruction efficiency we consider only jets which have been reconstructed within the cone $\Delta R = 0.4$ from the primary parton (particle) and we require that the primary parton (particle) has passed the same kinematical selection as required for reconstructed jets.

### 2.5.1 Labeling

Very important for the physics at LHC are jets originating from b-quarks (so called b-jets) which can be identified in the detector using b-tagging technique (vertex or soft-lepton tags). The package labels a jet as a b-jet if it is reconstructed within a limited rapidity range (default: $|\Delta\eta| < 2.5$) and if a b-quark of a transverse momenta (after FSR) above the threshold (default: $p_T > 5$ GeV) is found within the cone (default: $\Delta R = 0.2$) around the axis of reconstructed jet. The similar criteria are used for labeling the c-jets.

Equivalently important are also jets originating from the hadronic $\tau$-decay (so called $\tau$-jets) which can be identified using dedicated algorithms. The package labels jet as a tau-jet if the hadronic decay product is relatively hard (default: $p_T^{\tau-had} > 10$ GeV), inside limited rapidity range (default: $|\eta| < 2.5$), dominates reconstructed jet transverse momenta (default: $p_T^{\tau-had}/p_T^{jet} > 0.9$), and is within the cone (default: $\Delta R_{jet,\tau-had} < 0.3$) around the axis of a jet.

Table 1 summarises the jets reconstruction+labeling efficiencies as obtained for the $WH, H \to b\bar{b}, c\bar{c}, u\bar{u}$ and $gg \to H \to \tau\tau$ events. Jets labeling is optional, can be switched off for b- and c-jets and/or separately for tau-jets (default: ON).

Table 1: *Efficiency for jet reconstruction+labeling for different types of initial partons with $p_T^{parton} > 15$ GeV (required $p_T^{jet} > 15$ GeV). The rapidity coverage is limited to $|\eta| < 2.5$. The $\Delta R_{cone} = 0.4$ is used for cluster reconstruction and $\Delta R_{cone} = 0.2$ is used for matching criteria. The $gg \to H, H \to b\bar{b}, c\bar{c}, u\bar{u}$ and $gg \to H \to \tau\tau$ processes were generated with $m_H = 125$ GeV. In case of tau-jets only hadronic tau decays were generated.*

| Parton type | Reconstruction + Labeling |
|:---:|:---:|
| u-quark | 95% |
| b-quark | 81% |
| c-quark | 87% |
| tau-jet | 80% |

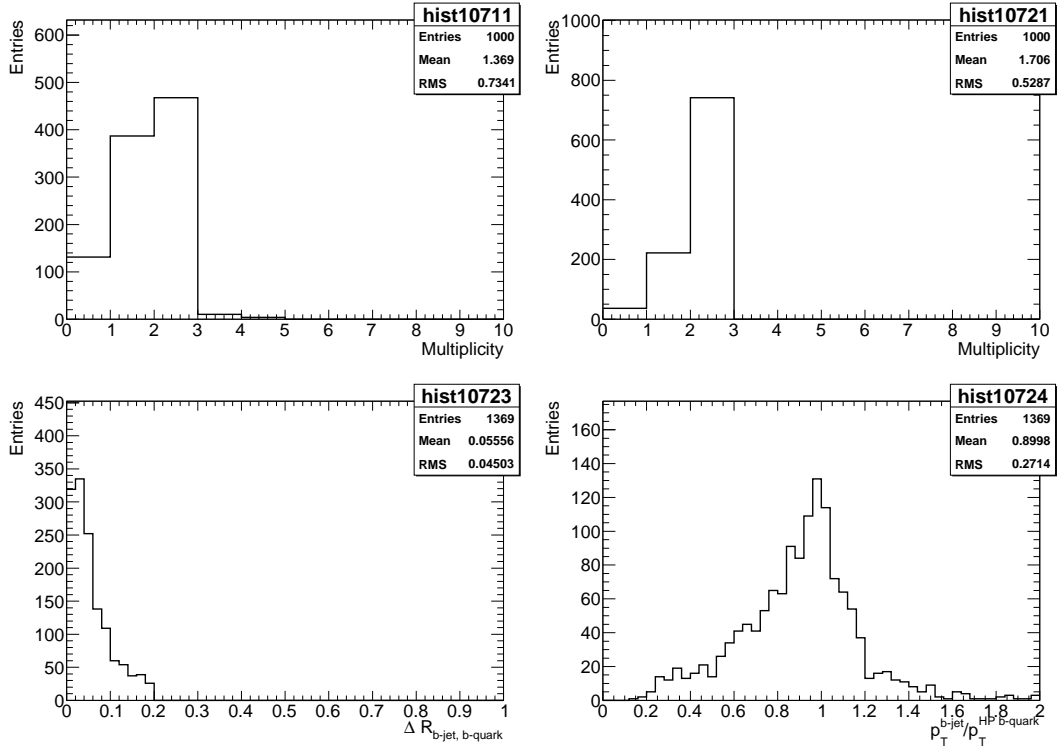

Figure 6: *Multiplicity of b-jets (top left) and hard-process b-quarks (top-right), $\Delta R$ cone distance between b-labeled jet and hard-process b-quark (bottom-left); the ratio of $p_T^{jet}/p_T^{HPb-quarks}$ (bottom-right). Distributions are shown for $gg \to H, H \to b\bar{b}$ and $m_H = 125~GeV$.*

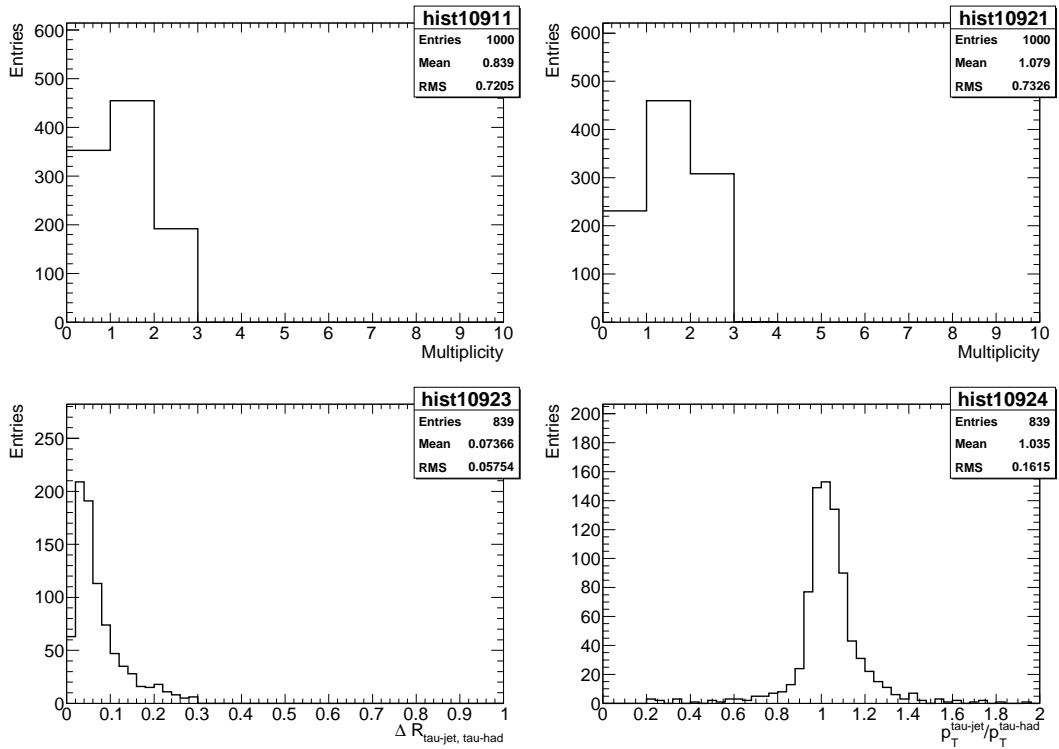

Figure 7: *Multiplicity of tau-jets (top left) and hadronic tau (top-right), $\Delta R$ cone distance between tau-labeled jet and hard-process tau-had (bottom-left); the ratio of $p_T^{tau-jet}/p_T^{tau-had}$ (bottom-right). Distributions are shown for $gg \to H, H \to \tau\tau$ and $m_H = 125$ GeV.*

### 2.5.2 Calibration

The reconstructed jets four-momenta need to be corrected for the out-cone energy loss (cascade outside the jet-cone) and for the loss of the particles escaping detection (those below threshold at $p_T = 0.5$ GeV, neutrinos, invisible particles, muons outside acceptance range or below the observability threshold). Such correction, called calibration, can be performed on the statistical basis only. The single default calibration function (the same for any type of jets), as a function of transverse momenta of reconstructed jet, $p_T^{jet}$, is provided in the package. The calibration algorithm corrects jets four-momenta without altering their direction. Calibration can be switched off (default: ON).

The quality of the calibration algorithm can be verified by monitoring the ratio of $p_T^{b-quark}/p_T^{b-jet}$, taking into account a hard-process quark which originates a given jet. One can note, (see Fig. 7, bottom plots) that the implemented calibration function has a tendency to undercalibrate b-jets while calibrates reasonably well u-jets. We can observe also rather large tail in the $p_T^{jet}/p_T^{b-quark}$ distribution, caused by the semileptonic b-quarks decays and larger spread of the cascading particles than in the case of light jets.

As an effect of the hadronisation and cascading decays, the expected resolution for the resonance reconstruction in the hadronic channels will be much worse than in the leptonic ones. The precision for reconstructing the peak position of the invariant mass of the di-jet system will relay on the precision of the calibration procedure.

## 2.6 Missing transverse energy

The missing transverse energy is calculated by summing up transverse momenta of identified isolated photons, electrons and muons, of jets and clusters not accepted as jets and of non-isolated muons not added to any jet. Finally, the transverse energies deposited in cells not used for clusters reconstruction are also included in the total sum. Transverse energies deposited in unused cells are smeared with the same energy resolution function as for jets, and cells with deposited transverse energy below a given threshold (default: 0 GeV) are excluded from the sum. From the calculation of the total sum $E_T^{obs}$ the missing transverse energy is obtained, $E_T^{miss} = -E_T^{obs}$ as well as the missing transverse momentum components $p_x^{miss} = -p_x^{obs}$, $p_y^{miss} = -p_y^{obs}$. The total calorimeter transverse energy, $\sum E_T^{calo}$, is calculated as the sum of all the above transverse energies except that of muons. Please note, that missing transverse energy is calculated from the energy balance before jets calibration is performed, as the possible out-cone energy loss is already taken into account by summing up energy deposition of unused cells/clusters.

Fig. 8 shows the resolution of the transverse missing energy obtained for the di-jet events generated with the transverse momenta of the hard process above 17 GeV.

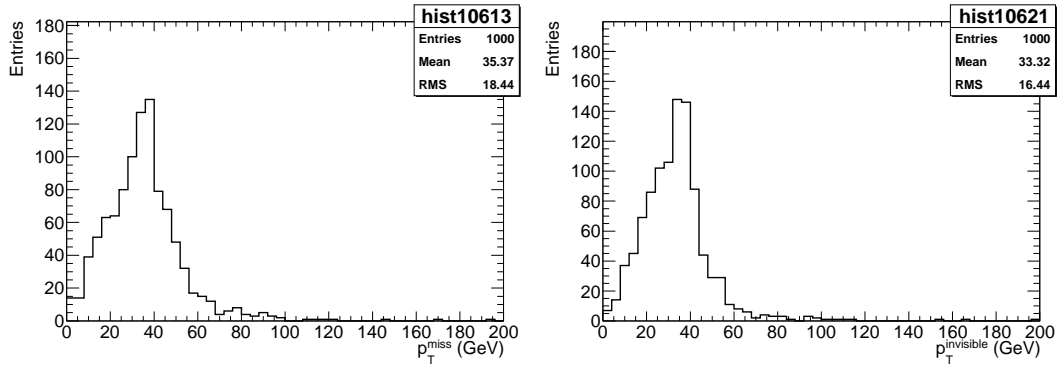

Figure 8: *The reconstructed missing transverse energy (left) and transverse momenta on neutrino in $W \to e\nu$ events.*

## 2.7 Additional efficiencies

The algorithms of the `AcerDET` package are not correcting for inefficiencies of photon, electron and muon reconstruction and identification. To be more realistic, one should apply a weighting factor of 70%-90% for each isolated lepton used in the analysis and of 80% for each isolated photon.

The package is also not correcting for tagging-efficiencies, namely the labeling procedure is not equivalent to the b- and tau-jet identification in the experiment. One can assume b-tagging efficiency of 60% per b-labeled jet with mistagging probability of 10% for c-labeled jet and 1% for the light jet. For the tau-jets using efficiency of 50% (per tau-labelled jet) and 5 - 10 % mistagging probability for other jets could be a reasonable assumptions. One should be well aware that what proposed above represents quite crude estimates.

The package is also not correcting for trigger efficiencies which has to be applied in addition if a given reconstructed object is foreseen to trigger an event.

### 2.8 OUTPUT format

By the end of the simulation and reconstruction algorithm, reconstructed entities: photons, electrons, muons, jets, transverse missing energy are available as collections of objects in the `OutputRecord`. Provided is algorithm to store this information in the `ROOT` [11] tree. However, the user might decide to use his preferred data-base technology for storing output from the simulation and reconstruction algorithms.

## 3 Outlook

We presented an update from `Fortran` to `C++` of a package which can be useful for several phenomenological studies on the high $p_T$ physics at LHC. As examples of applications some recent publications which have been using its fortran version AcerDET-1.0 [1] are papers eg. [14, 15, 16, 17] ( for the more complete list see link [2] ).

It is not the aim of the package to represent in details performance neither of ATLAS nor of CMS detectors, nevertheless some global features of these would be reproduced well. We have presented some benchmark results in Section 2. In particular we believe that the analyses performed with the package for physics at LHC will be more realistic than parton-level studies alone.

The package allows also for flexible adjusting of key parameters which characterise features of any detector be it for LHC or for future ILC or FFC experiments.

## Acknowledgments

This work on AcerDET-1.0 was inspired by the several years of ERW involvement in the activity of the Physics Working Groups of the ATLAS Collaboration. ERW grateful to all colleagues for a very creative atmosphere. In particular, for several suggestions and inspiring discussions to, Daniel Froidevaux, who some years ago initiated and guided her work on the first version of the fast simulation package. The AcerDET-2.0 represents continuation of this project in new software environment.

The work of P. Mikos and E. Richter-Was is supported in part by the Polish National Centre of Science, Grant No. DEC-2011/03/B/ST2/00220. Work of E. Richter-Was is also supported in part by the Executive Research Agency (REA) of the European Commission under the Grant Agreement PITN-GA-2012-316704 (HiggsTools).

# A General informations

The simulation and reconstruction algorithm is executed by a call to the routine `AcerDET.analyseRecord`. The events which is going to be processed should be stored in the `InputRecord`.

The convention for the particles status, mother-daughter relations, particles codes, etc. should be the same as in the `HepMC` standard.

The input/output logical identifiers should be defined in `configFileName`.

The package reads single input file `acerdet.dat` which contains parameters for events simulation and reconstruction and write out single control output file `acerdet.out`. If initialised by the user in the main program, control histograms and ntuple with reconstructed events will be stored in the `XXX.root` file, where `XXX` stands for file which configures generator used for generating events. In the distributed example the `Pythia 8.1` is used.

## A.1 Class AcerDET

This is the main class, called to execute simulation and reconstruction algorithms. The following methods are implemented:

- `AcerDET contructor` – initalisation, which should be called before the first event is processed.

- `AcerDET::analyseRecord` – simulation and reconstruction, should be called event by event.

- `AcerDET destructor` – finalisation, should be called after last event is processed.

The `AcerDET::analyseRecord` invokes other methods: `analyse_Cell`, `analyse_Cluster`, `analyse_Muon`, `analyse_Electron`, `analyse_Photon`, `analyse_Jet`, `analyse_Mis`, `analyse_BJet`, `analyse_CJet`, `analyse_Tau`, `analyse_Calibration`. The order in which methods are invoked should not be changed.

The `AcerDET::printInfo()` invokes priting information on the configurations used for individual methods and `AcerDET::printResults()` summary information on the recontructed objects.

## A.2 Interfaces to Event Generators

The event, which is going to be processed, should be stored in the `HepMC` format. In the source code of the program provided are methods to traslate this format into internal event structure, vector of `Particles` stored in `InputRecord` and to interpret status and pdgId codes of the particles.

The following internal types are introduced, based on the information available in the `HepMC` conversion of `Pythia 8.1` event record.

- ParticleType

  - PT_CJET - particle with pdgId code = -4, 4
  - PT_BJET - particle with pdgId code = -5, 5
  - PT_ELECTRON - particle with pdgId code = -11, 11

- PT_NEUTRINO_ELE - particle with pdgId code = -12, 12
- PT_MUON - particle with pdgId code = -13, 13
- PT_NEUTRINO_MUON - particle with pdgId code = -14, 14
- PT_TAU - particle with pdgId code = -15, 15
- PT_NEUTRINO_TAU - particle with pdgId code = -16, 16
- PT_PHOTON - particle with pdgId code = 22
- PT_BOSON_Z - particle with pdgId code = 23
- PT_BOSON_W - particle with pdgId code = -24, 24
- PT_BOSON_H - particle with pdgId code = 25
- PT_UNKNOWN - other

- `ParticleStatus`

  - PS_BEAM - corresponding HepMC: part->is_beam() = true;
  - PS_FINAL - corresponding HepMC: part->is_undecayed() = true;
  - PS_DECAYED - corresponding HepMC: part->has_undecayed() = true;
  - PS_HISTORY - corresponding HepMC: part->is_undecayed() = true;
  - PS_CASCADE_QUARK - corresponding HepMC: abs(part->status()) >=30;
  - PS_HP_QUARK - corresponding HepMC: abs(part->status()) = 20 - 29;
  - PS_NULL - other;

While `ParticleStatus` and `ParticleType` become an attribute of the object `Particle`, the method `AcerDet::core::isHardProcess` is also provided to decide if a given particle is from hard process. This method (so far) relies on information of the particle origin, i.e. type of the mother particle, but more refined procedure can be impmeneted there as well.

### A.3 External calling sequence

The following calling sequence should be provided in the main program (see `demo.cpp`):

- initialise generator
- initialise and configure `AcerDET`
- initialise root tree and histogram manager
- start even loop
  - generate new event
  - convert to `HepMC` format (if needed)
  - process with `AcerDET`
  - fill root tree with reconstructed objets
- write tree and histograms into file, write information into log file.

Invoking method which fill root tree is optional. User however may decide to skip this part, or implement different output format.

In general, if user do not wish, the dependencies on `ROOT` could be easily removed. It requires linking into executable the respective alternative library and providing very simple conversion package where the existing calls to `ROOT` functions for histograming are used to fill information into the user's preferred ones.

### A.4   Execution sequence

After the initialisation phase, the event by event execution sequence is a following one:

- Map of the cells energy deposition is created: `analyse_Cell`.

- Calorimetric clusters are reconstructed: `analyse_Cluster`.

- Isolated and non-isolated muons are reconstructed: `analyse_Muon`.

- Isolated electrons are reconstructed: `analyse_Electron`.

- Isolated photons are reconstructed: `analyse_Photon`.

- The remaining calorimetric clusters are identified as jets: `analyse_Jet`.

- Missing transverse is calculated for the reconstructed event: `analyse_Mis`.

- Optionally, if specified in the acerdet.dat file, algorithms for jets labeling and calibration of jets are executed (default=ON): `analyse_BJet, analyse_CJet, analyse_Tau, analyse_Calibration`

- Reconstructed event is stored in the final record: `OutputRecord`.

The respective sequence of calls is executed in the `AcerDET::analyseRecord`.

### A.5   Structure of the distribution version

The distribution version consists of the source code of the `AcerDET` and example of the main program for execution with `PYTHIA 8.1` generator.

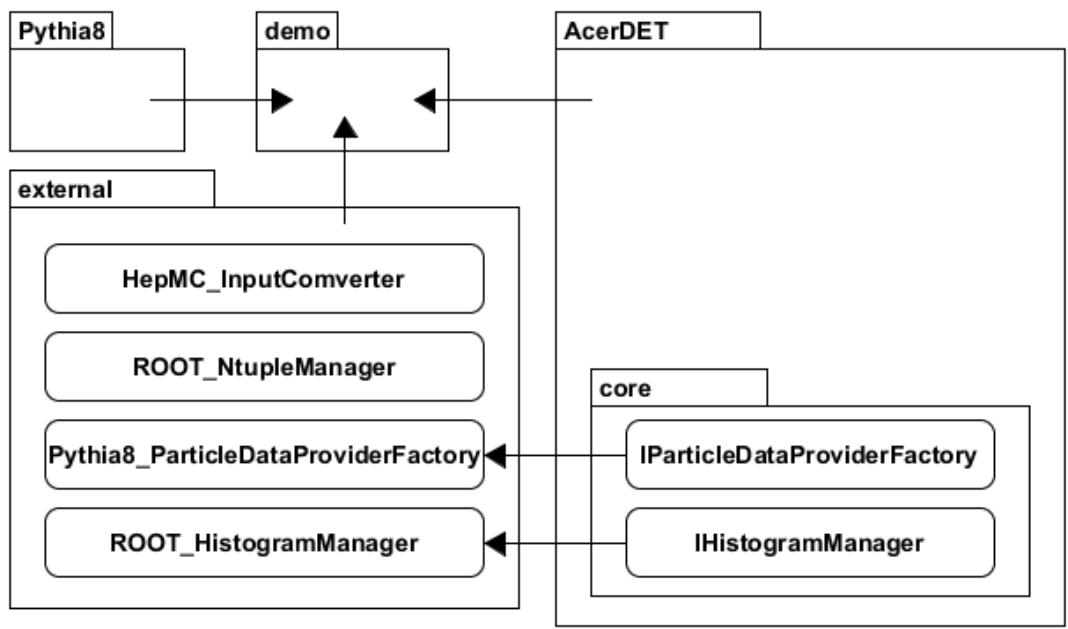

Figure 9: *Structure of AcerDET package.*

AcerDET is designed to be an universal tool. It does not depend on any external library (except `C++ STL`). In order to provide functionalities not built-in directly in `AcerDET`, it provides a set of interfaces.

Instantiation of `AcerDET` requires providing instances of such interfaces. Because of separation between analyse logic and data structures `AcerDET` might work with a various number of external libraries.

The `AcerDET` package provides a set of default implementations for its interfaces, grouped in `AcerDET::external` subpackage containing:

- `Root_HistogramManager` – a default implementation of `IHistogramManager` interface. Provides methods for storing weighted data in named histograms. For more details see `AcerDET::core::IHistogramManager`.

- `Pythia8_ParticleDataProviderFactory` – a default implementation of `IParticleDataProviderFactory` interface. Represents a particle types database, providing particle charge, name, etc. For more details see `AcerDET::core::ParticleDataProvider`.

- `HepMC_InputConverter` – static class converting `HepMC` event record to `AcerDET` internal representation of input record. For more details see `AcerDET::io::InputRecord`.

- `Root_NTupleManager` – wrapper for `ROOT`'s NTuple. Provides methods for storing and saving weighted events. The `demo` uses it like an external database.

The `acerdet.dat` file (configuration file for `AcerDET` package) resides in main directory `acerdet_dat`. Source code resides in `src` which has one level deeper internal structure (see

README there). Interfaces to external libraries are in directory `external` The library will be created in `src` directory, provided is also example of the main program `demo.cpp` to execute `AcerDET` algorithms with `PYTHIA 8.1` generator:

- Struture of directory **acerdet**

```
-rw-rw-r--  1 erichter erichter     3703 Jun 10 16:07 acerdet.dat
-rw-rw-r--  1 erichter erichter       65 Jun 10 16:07 common.inc
drwxrwxr-x  2 erichter erichter     4096 Jun 12 10:31 conf
-rw-rw-r--  1 erichter erichter     2719 Jun 10 17:25 demo.cpp
-rw-rw-r--  1 erichter erichter      590 Jun 10 16:07 demoCreateConfig.cpp
drwxrwxr-x  3 erichter erichter     4096 Jun 12 10:31 external
-rw-rw-r--  1 erichter erichter      861 Jun 10 16:08 external.inc
drwxrwxr-x  8 erichter erichter     4096 Jun 12 17:55 .git
-rw-rw-r--  1 erichter erichter       74 Jun 10 16:09 .gitignore
-rw-rw-r--  1 erichter erichter      848 Jun 10 16:07 Makefile
-rw-rw-r--  1 erichter erichter      204 Jun 10 16:08 path.inc
-rw-rw-r--  1 erichter erichter      274 Jun 10 16:07 README
-rw-rw-r--  1 erichter erichter       69 Jun 10 16:07 setup.sh
drwxrwxr-x  6 erichter erichter     4096 Jun 12 10:31 src
```

- Struture of directory **acerdet/src** (after compilation using 'make')

```
-rw-rw-r-- 1 erichter erichter     4078 Jun 10 17:25 AcerDET.cpp
-rw-rw-r-- 1 erichter erichter     3994 Jun 10 17:25 AcerDET.h
-rw-rw-r-- 1 erichter erichter   264412 Jun 10 17:25 AcerDET.o
drwxrwxr-x 2 erichter erichter     4096 Jun 12 10:31 analyse
drwxrwxr-x 2 erichter erichter     4096 Jun 10 17:25 conf
drwxrwxr-x 2 erichter erichter     4096 Jun 11 19:14 core
-rw-rw-r-- 1 erichter erichter    99229 Jun 10 16:08 doxygen.config
drwxrwxr-x 2 erichter erichter     4096 Jun 10 17:25 io
-rw-rw-r-- 1 erichter erichter  5538158 Jun 12 10:31 libAcerDET.a
-rwxrwxr-x 1 erichter erichter  3803523 Jun 12 10:31 libAcerDET.so
-rw-rw-r-- 1 erichter erichter     1689 Jun 10 16:07 Makefile
-rw-rw-r-- 1 erichter erichter      132 Jun 10 16:08 path.inc
-rw-rw-r-- 1 erichter erichter      252 Jun 10 16:07 README
```

To execute the package following actions should be taken:

- In subdirectory `acerdet` edit files `external.inc` and path.inc. Then type: *make* . It will compile the code and create the `src/libAcerDET.a` and `src/libAcerDET.so`. It will create also demo executable `demo.exe`

- To execute, type
  `./demo.exe conf/XXX.conf nEvent`
  where `conf/XXX.conf` stands for file configuring generated process and `nEvent` is a integer number of required events to be generated.

- Output will be stored in file `conf/XXX.conf.root`

### A.6 Development

Regarding modifications in `AcerDET` algorithms, package users are able to develop own extensions to our main algorithm. As it was shown in previous section `AcerDET` provides a set of interfaces to implement in order to use the package. `AcerDET::external` sub-package contains default implementations of those interfaces, but user may develop his own implementations. For instance if someone wants to use other histogram manager, it is enough to create new class inheriting from `AcerDET::core::IHistogramManager` and pass it's instance when create new `AcerDET` instance (as it is shown in `demo.cpp`).

```
const std::string configFileName = "acerdet.dat";
Configuration configuration = Configuration::fromFile( configFileName );
IParticleDataProviderFactory *pdpFactory =
new external::Pythia8_ParticleDataProviderFactory();
IHistogramManager *histoManager =
new external::Root_HistogramManager(); // HERE USE OWN INSTANCE
external::Root_NTupleManager nTuple;

AcerDET acerDet(
configuration,
pdpFactory,
histoManager
);
```

Analoguously user may use own instance of `IParticleDataProviderFactory`. ROOT NTuple is used in demo like na external database, so using other databases is possible in similiar way. Distributed demo version uses `HepMC` as input data standard, while `AcerDET` uses its own input format. Default conversion algoritm is placed in `AcerDET::external::HepMC_InputConverter`. In order to use other input format, similiar converter to `AcerDET` input format is required.

# B  List of input parameters: `acerdet.dat` file

```
Flag.HistogramID            10000        ....id for histograms
Flag.Smearing               1            ....smearing on=1, off=0
Flag.B-Field                1            ....B-field   on=1, off=0
Flag.SUSY_LSP_Particle      66           ....code for SUSY LSP particle
Flag.BC-JetsLabeling        1            ....b- and c-jets labeling on=1, off=0
Flag.Tau-JetsLabeling       1            ....tau-jets labeling on=1, off=0
Flag.JetCalibration         1            ....jet calibration  on=1, off=0

Cell.RapidityCoverage       5.000        ....rapidity coverage
Cell.MinpT                  0.500        ....min p_T for B-field
Cell.MinEt                  0.000        ....min E_T for cell
Cell.EtaTransition          3.200        ....eta transition in cells granularity
Cell.GranularityEta         0.100        ....granularity in eta (within Cell.EtaTransition), 2x outside
Cell.GranularityPhi         0.100        ....granularity in phi (within Cell.EtaTransition), 2x outside

Cluster.MinEt               5.000        ....minimum E_T for cluster
Cluster.ConeR               0.400        ....cone R for clustering
Cluster.RapidityCoverage    5.000        ....rapidity coverage
Cluster.MinEtInit           1.500        ....min E_T for cluster initiator

Muon.MinMomenta             6.000        ....minimum muon-momenta to be detected
Muon.MaxEta                 2.500        ....maximum muon eta to be detected
Muon.MinIsolRlj             0.400        ....min R_lj for muon-isolation
Muon.ConeR                  0.200        ....R_cone for energy deposition
Muon.MaxEnergy              10.000       ....max energy deposition for isol

Photon.MinMomenta           5.000        ....minimum photon-momenta to be isol
Photon.MaxEta               2.500        ....maximum photon eta to be isol
Photon.MinJetsRlj           0.150        ....min R_lj for photon-jet
Photon.MinIsolRlj           0.400        ....min R_lj for photon-isolation
Photon.ConeR                0.200        ....R_cone for energy deposition
Photon.MaxEnergy            10.000       ....max energy deposition for isol

Electron.MinMomenta         5.000        ....minimum electron-momenta to be isol
Electron.MaxEta             2.500        ....maximum electron eta to be isol
Electron.MinJetsRlj         0.150        ....min R_lj for electron-jet
Electron.MinIsolRlj         0.400        ....min R_lj for electron-isolation
Electron.ConeR              0.200        ....R_cone for energy deposition
Electron.MaxEnergy          10.000       ....max energy deposition for isol

Jets.MinEnergy              10.000       ....jets energy_min  threshold
Jets.RapidityCoverage       5.000        ....rapidity coverage for jets

BJets.MinMomenta            5.000        ....minimum b-quark pT (after FSR) momenta for b-jet label
BJets.MaxEta                2.500        ....maximum b-quark eta for  b-jet label
BJets.MaxRbj                0.200        ....max R_bj for b-jet label

CJets.MinMomenta            5.000        ....minimum c-quark pT (after FSR) momenta for c-jet label
CJets.MaxEta                2.500        ....maximum c-quark eta for  c-jet label
CJets.MaxRcj                0.200        ....max R_cj for c-jet label

Tau.MinpT                   10.000       ....minimum tau-had pT for  tau-jet label
Tau.MaxEta                  2.500        ....maximum tau-eta for  tau-jet label
Tau.MinR                    0.300        ....max R_tauj for tau-jet
Tau.MaxR                    0.900        ....max R_tauj for tau-jet

Misc.MinEt                  0.000        ....min E_T for energy in cell to count unused cell
```

# C Reconstructed entities

Reconstructed entities are stored as vectors of `ObjectData` (for isolated photon, muons, electrons), as `CellData` (for cells), as `JetData` (for jets) and `MisData` for missing energy. Objects in each category of `ObjectData` and `JetData` are ordered with the decreasing transverse momenta.

## C.1 ObjectData

For `ObjectData` stored is information of the `pdg_id`, so the information of the charge is preserved, `eta`, `phi` and `pT` which allows to recalculate full kinematics assuming that mass is equal to zero, and the internal flag `alreadyUsed` needed for calculating missing tranverse energy for the event.

## C.2 JetData

For `JetData` stored is information on the `type` which flags if a given jet is b-quark tagged, c-quark tagged or matched with hadronic-tau decays. The angular kinematic variables `eta`, `phi` and `eta_rec`, `phi_rec` and `pT`.

## C.3 MisData

For `MisData` stored is information on reconstructed momenta both transverse components of electrons, muons, photons, clusters and jets, `PXSUM, PYSUM`, on reconstructed total tranverse components i.e. after adding non-clusterd cells `PXREC, PYREC`, sum of transverse momenta components, the reconstructed transverse momenta `SUMET`, the transverse components of momenta of neutrinos and invisible particles (as defined by the user) in the event `PXNUES, PYNUES` and finally sum of the total transverse energy components deposited in the calorimeter `PXXCALO, PYYCALO`.

# D Content of the `ACDTree`

Reconstructed objects and, for convenience of the analysis, also partial information on generated event are stored in the format of `ROOT` tree, the `ACDTree`.

- Generator info: Event weight and identifier for generated process.

- Generated particles: stored are those participating in the *hard-scattering*, which is defined using `isHardProcess` method and flags: `PS_Beam`, `PS_HISTORY`. Stored are 4-momenta, and `pdg_id` code.

- Isolated photons, electrons, muons: stored are 4-momenta, and `pdg_id` code.

- Jets: stored are 4-momenta, the `pdg_id` code indicate if jet was labeled as -batteg, c-tagged or hadronic-decay of tau lepton.

- Missing energy: stored are transverse momenta components of missing energy, invisible energy and energy reconstructed in the calorimeter.

Additional information, if required, can be also extracted from the internal variables of simulation and reconstruction algorithms: eg. on used/unused cells, non-isolated muons, unused clusters, etc. and added to the `ACDTree`.

# E  Parametrisation for energy/momenta resolution

## E.1  Smearing::forPhoton

The parametrisation for photon energy resolution assumes only energy dependence and the Gaussian smearing with $10\%/\sqrt{E_\gamma}$ resolution.

$$E_\gamma^{smeared} = E_\gamma^{true} \cdot (1 + r_n \cdot \frac{0.10}{\sqrt{E_\gamma^{true}}}). \qquad (1)$$

Where $r_n$ is the random number generated according to the Gaussian distribution (normal distribution from Standard C++ library). All 4-momenta components of the photon are smeared with the same resolution so the direction of the photon is not altered.

## E.2  Smearing::forElectron

The parametrisation for electron energy resolution assumes only energy dependence and the Gaussian smearing with $12\%/\sqrt{E_e}$ resolution.

$$E_e^{smeared} = E_e^{true} \cdot (1 + r_n \cdot \frac{0.12}{\sqrt{E_e^{true}}}). \qquad (2)$$

Where $r_n$ is the random number generated according to the Gaussian distribution (normal distribution from Standard C++ library). All 4-momenta components of the electron are smeared with the same resolution so the direction of the electron is not altered.

## E.3  Smearing::forHadron

The parametrisation for clusters momenta resolution assumes only energy dependence and Gaussian smearing with $50\%/\sqrt{E_{clu}}$ or $100\%/\sqrt{E_{clu}}$ resolution. The transition region, $|\eta^{clu}| = CALOTH$ is the same as the transition of the cells granularity.
For $|\eta^{clu}| < CALOTH$:

$$E_{clu}^{smeared} = E_{clu}^{true} \cdot (1 + r_n \cdot \frac{0.50}{\sqrt{E_{clu}^{true}}}). \qquad (3)$$

For $|\eta^{clu}| > CALOTH$:

$$E_{clu}^{smeared} = E_{clu}^{true} \cdot (1 + r_n \cdot \frac{1.00}{\sqrt{E_{clu}^{true}}}). \qquad (4)$$

Where $r_n$ is the random number generated according to the Gaussian distribution (normal distribution from Standard C++ library). All 4-momenta components of the muon are smeared with the same resolution so the direction of the muon is not altered.

### E.4 Smearing::forMuon

The parametrisation for muon momenta resolution assumes only transverse momenta dependence and Gaussian smearing with $0.0005 \cdot pT_\mu$ resolution.

$$pT_\mu^{smeared} = pT_\mu^{true}/(1 + r_n \cdot 0.0005 \cdot pT_\mu^{true}). \tag{5}$$

Where $r_n$ is the random number generated according to the Gaussian distribution (normal distribution from Standard C++ library). All 4-momenta components of the muon are smeared with the same resolution so the direction of the muon is not altered.

### E.5 Particle::foldPhi

The effect of the magnetic field is included by simple shifting the $\phi$ position of the charged particles respectively to its transverse momenta, parametrised as following:

$$\delta\phi = 0.5/pT^{part}. \tag{6}$$

The $\delta\phi$ is calculated in radians and $pT^{part}$ is given in $GeV$. The sign of the $\delta\phi$ is the same as the sign of the particle charge. Charged particles with $pT^{part} < 0.5$ GeV are assumed to be looping in the detector and not depositing energy in the calorimeter.

## F  Some formulas

We work with the assumptions of massless reconstructed objets. The following relations were used to translate between $(p_T, \eta, \phi)$ coordinates and four-momenta $(p_x, p_y, p_z, E)$.

$$p_x = p_T \cdot cos(\phi) \tag{7}$$
$$p_y = p_T \cdot sin(\phi) \tag{8}$$
$$p_z = p_T \cdot cosh(\eta) \tag{9}$$
$$E = p_T \cdot sinh(\eta) \tag{10}$$

$$p_T = \sqrt{p_x^2 + p_y^2} \tag{11}$$

$$\eta = sign(\ln \frac{\sqrt{p_T^2 + p_z^2} + |p_z|}{p_T}, p_z) \tag{12}$$

$$\phi = asinh(p_y/\sqrt{p_x^2 + p_y^2}) \tag{13}$$

# G Output content

## G.1 Output ACDTree content: `acerdet.ntup` file

The content of the ntuple is exactly as described in the previous section. As an auxiliary info, within the stored information, could be provided code on the generated process, `IDPROC`.

```
******************************************************************************
*Tree    :ACDTree   : ACDTree                                                 *
*Entries :  100000 : Total =       176704219 bytes  File  Size =    83438909 *
*        :         : Tree compression factor =    2.12                       *
******************************************************************************
*Br    0 :ProcessID : ProcessID/I                                             *
*Entries :  100000 : Total  Size=      785597 bytes  File Size  =      401027 *
*Baskets :      13 : Basket Size=       32000 bytes  Compression=   1.96     *
*............................................................................*
*Br    1 :part_n    : part_n/I                                                *
*Entries :  100000 : Total  Size=      785546 bytes  File Size  =      400988 *
*Baskets :      13 : Basket Size=       32000 bytes  Compression=   1.96     *
*............................................................................*
*Br    2 :part_pdgId : vector<int>                                            *
*Entries :  100000 : Total  Size=    14579602 bytes  File Size  =     7086584 *
*Baskets :     235 : Basket Size=       32000 bytes  Compression=   2.06     *
*............................................................................*
*Br    3 :part_mother_pdgId : vector<int>                                     *
*Entries :  100000 : Total  Size=    14581275 bytes  File Size  =     7088229 *
*Baskets :     235 : Basket Size=       32000 bytes  Compression=   2.06     *
*............................................................................*
*Br    4 :part_px   : vector<float>                                           *
*Entries :  100000 : Total  Size=    14578885 bytes  File Size  =     7085879 *
*Baskets :     235 : Basket Size=       32000 bytes  Compression=   2.06     *
*............................................................................*
*Br    5 :part_py   : vector<float>                                           *
*Entries :  100000 : Total  Size=    14578885 bytes  File Size  =     7085879 *
*Baskets :     235 : Basket Size=       32000 bytes  Compression=   2.06     *
*............................................................................*
*Br    6 :part_pz   : vector<float>                                           *
*Entries :  100000 : Total  Size=    14578885 bytes  File Size  =     7085879 *
*Baskets :     235 : Basket Size=       32000 bytes  Compression=   2.06     *
*............................................................................*
*Br    7 :part_E    : vector<float>                                           *
*Entries :  100000 : Total  Size=    14578646 bytes  File Size  =     7085644 *
*Baskets :     235 : Basket Size=       32000 bytes  Compression=   2.06     *
*............................................................................*
*Br    8 :ele_n     : ele_n/I                                                 *
*Entries :  100000 : Total  Size=      785529 bytes  File Size  =      400975 *
*Baskets :      13 : Basket Size=       32000 bytes  Compression=   1.96     *
*............................................................................*
*Br    9 :ele_pdgId : vector<int>                                             *
*Entries :  100000 : Total  Size=     3204969 bytes  File Size  =     1411515 *
*Baskets :      57 : Basket Size=       32000 bytes  Compression=   2.27     *
*............................................................................*
*Br   10 :ele_px    : vector<float>                                           *
*Entries :  100000 : Total  Size=     3204786 bytes  File Size  =     1411344 *
*Baskets :      57 : Basket Size=       32000 bytes  Compression=   2.27     *
*............................................................................*
*Br   11 :ele_py    : vector<float>                                           *
*Entries :  100000 : Total  Size=     3204786 bytes  File Size  =     1411344 *
*Baskets :      57 : Basket Size=       32000 bytes  Compression=   2.27     *
*............................................................................*
*Br   12 :ele_pz    : vector<float>                                           *
*Entries :  100000 : Total  Size=     3204786 bytes  File Size  =     1411344 *
*Baskets :      57 : Basket Size=       32000 bytes  Compression=   2.27     *
```

```
*............................................................................*
*Br   13 :ele_E     : vector<float>                                            *
*Entries :   100000 : Total  Size=    3204725 bytes  File Size  =    1411287 *
*Baskets :       57 : Basket Size=      32000 bytes  Compression=   2.27     *
*............................................................................*
*Br   14 :muo_n     : muo_n/I                                                  *
*Entries :   100000 : Total  Size=     785529 bytes  File Size  =     400975 *
*Baskets :       13 : Basket Size=      32000 bytes  Compression=   1.96     *
*............................................................................*
*Br   15 :muo_pdgId : vector<int>                                              *
*Entries :   100000 : Total  Size=    3204997 bytes  File Size  =    1411543 *
*Baskets :       57 : Basket Size=      32000 bytes  Compression=   2.27     *
*............................................................................*
*Br   16 :muo_px    : vector<float>                                            *
*Entries :   100000 : Total  Size=    3204814 bytes  File Size  =    1411372 *
*Baskets :       57 : Basket Size=      32000 bytes  Compression=   2.27     *
*............................................................................*
*Br   17 :muo_py    : vector<float>                                            *
*Entries :   100000 : Total  Size=    3204814 bytes  File Size  =    1411372 *
*Baskets :       57 : Basket Size=      32000 bytes  Compression=   2.27     *
*............................................................................*
*Br   18 :muo_pz    : vector<float>                                            *
*Entries :   100000 : Total  Size=    3204814 bytes  File Size  =    1411372 *
*Baskets :       57 : Basket Size=      32000 bytes  Compression=   2.27     *
*............................................................................*
*Br   19 :muo_E     : vector<float>                                            *
*Entries :   100000 : Total  Size=    3204753 bytes  File Size  =    1411315 *
*Baskets :       57 : Basket Size=      32000 bytes  Compression=   2.27     *
*............................................................................*
*Br   20 :pho_n     : pho_n/I                                                  *
*Entries :   100000 : Total  Size=     785529 bytes  File Size  =     400975 *
*Baskets :       13 : Basket Size=      32000 bytes  Compression=   1.96     *
*............................................................................*
*Br   21 :pho_pdgId : vector<int>                                              *
*Entries :   100000 : Total  Size=    4558940 bytes  File Size  =    2093066 *
*Baskets :       78 : Basket Size=      32000 bytes  Compression=   2.18     *
*............................................................................*
*Br   22 :pho_px    : vector<float>                                            *
*Entries :   100000 : Total  Size=    4558694 bytes  File Size  =    2092832 *
*Baskets :       78 : Basket Size=      32000 bytes  Compression=   2.18     *
*............................................................................*
*Br   23 :pho_py    : vector<float>                                            *
*Entries :   100000 : Total  Size=    4558694 bytes  File Size  =    2092832 *
*Baskets :       78 : Basket Size=      32000 bytes  Compression=   2.18     *
*............................................................................*
*Br   24 :pho_pz    : vector<float>                                            *
*Entries :   100000 : Total  Size=    4558694 bytes  File Size  =    2092832 *
*Baskets :       78 : Basket Size=      32000 bytes  Compression=   2.18     *
*............................................................................*
*Br   25 :pho_E     : vector<float>                                            *
*Entries :   100000 : Total  Size=    4558612 bytes  File Size  =    2092754 *
*Baskets :       78 : Basket Size=      32000 bytes  Compression=   2.18     *
*............................................................................*
*Br   26 :jet_n     : jet_n/I                                                  *
*Entries :   100000 : Total  Size=     785529 bytes  File Size  =     400975 *
*Baskets :       13 : Basket Size=      32000 bytes  Compression=   1.96     *
*............................................................................*
*Br   27 :jet_pdgId : vector<int>                                              *
*Entries :   100000 : Total  Size=    4991805 bytes  File Size  =    2301791 *
*Baskets :       85 : Basket Size=      32000 bytes  Compression=   2.17     *
*............................................................................*
*Br   28 :jet_px    : vector<float>                                            *
*Entries :   100000 : Total  Size=    4991538 bytes  File Size  =    2301536 *
*Baskets :       85 : Basket Size=      32000 bytes  Compression=   2.17     *
*............................................................................*
```

```
*Br   29 :jet_py    : vector<float>                                    *
*Entries :   100000 : Total  Size=    4991538 bytes  File Size  =    2301536 *
*Baskets :       85 : Basket Size=      32000 bytes  Compression=   2.17     *
*............................................................................*
*Br   30 :jet_pz    : vector<float>                                    *
*Entries :   100000 : Total  Size=    4991538 bytes  File Size  =    2301536 *
*Baskets :       85 : Basket Size=      32000 bytes  Compression=   2.17     *
*............................................................................*
*Br   31 :jet_E     : vector<float>                                    *
*Entries :   100000 : Total  Size=    4991449 bytes  File Size  =    2301451 *
*Baskets :       85 : Basket Size=      32000 bytes  Compression=   2.17     *
*............................................................................*
*Br   32 :pxmiss    : pxmiss/F                                         *
*Entries :   100000 : Total  Size=     785546 bytes  File Size  =     400988 *
*Baskets :       13 : Basket Size=      32000 bytes  Compression=   1.96     *
*............................................................................*
*Br   33 :pymiss    : pymiss/F                                         *
*Entries :   100000 : Total  Size=     785546 bytes  File Size  =     400988 *
*Baskets :       13 : Basket Size=      32000 bytes  Compression=   1.96     *
*............................................................................*
*Br   34 :pxnue     : pxnue/F                                          *
*Entries :   100000 : Total  Size=     785529 bytes  File Size  =     400975 *
*Baskets :       13 : Basket Size=      32000 bytes  Compression=   1.96     *
*............................................................................*
*Br   35 :pynue     : pynue/F                                          *
*Entries :   100000 : Total  Size=     785529 bytes  File Size  =     400975 *
*Baskets :       13 : Basket Size=      32000 bytes  Compression=   1.96     *
*............................................................................*
*Br   36 :pxcalo    : pxcalo/F                                         *
*Entries :   100000 : Total  Size=     785546 bytes  File Size  =     400988 *
*Baskets :       13 : Basket Size=      32000 bytes  Compression=   1.96     *
*............................................................................*
*Br   37 :pycalo    : pycalo/F                                         *
*Entries :   100000 : Total  Size=     785546 bytes  File Size  =     400988 *
*Baskets :       13 : Basket Size=      32000 bytes  Compression=   1.96     *
*............................................................................*
r
```

## G.2 Control printout

```
************************************************************
*                   AcerDET, version: 2.0                  *
*                   Released at: 30.06.2015                *
*                                                          *
*  Simplied event simulation and reconstruction package    *
*                                                          *
*          by E. Richter-Was (Institute of Physics)        *
*          and P. Mikos (Theoretical Computer Science)     *
*          Jagiellonian University, Cracow, Poland         *
************************************************************

 Initial configuration:
**************************************
*                                    *
*     *************************      *
*     ***   analyse::Cell    ***     *
*     *************************      *
*                                    *
**************************************

 clusters definition ...
 eta coverage  5.000000
 E_T_min cell thresh 0.000000
 eta gran. transition 3.200000
 gran in eta(central) 0.100000
 gran in phi(central) 0.100000

 B field apply ....
 B-field on
 p_T min non looping 0.500000

**************************************
*                                    *
*     *************************      *
*     ***   analyse::Cluster ***     *
*     *************************      *
*                                    *
**************************************

 clusters definition ...
 E_T_min cluster 5.000000
 E_T_min cell initia 1.500000
 R cone 0.400000
 eta coverage 5.000000
 eta gran. transition 3.200000

**************************************
*                                    *
*     *************************      *
*     ***   analyse::Muon    ***     *
*     *************************      *
*                                    *
**************************************

... muon isolation ...
min. muon p_T 6.000000
max. muon eta 2.500000
min R_lj for isolation 0.400000
R for energy deposit 0.200000
max E_dep for isolation 10.000000
smearing on
```

```
****************************************
*                                      *
*      ***************************     *
*      ***    analyse::Electron  ***   *
*      ***************************     *
*                                      *
****************************************

 ... electron isolation ...
 min. lepton p_T 5.000000
 max. lepton eta 2.500000
 max R_ej for ele-clu 0.150000
 min R_lj for isolation 0.400000
 R for energy deposit 0.200000
 max E_dep for isolation 10.000000
 smearing on

****************************************
*                                      *
*      **************************      *
*      ***    analyse::Photon   ***    *
*      **************************      *
*                                      *
****************************************

 ... photon isolation ...
 min. photon p_T 5.000000
 max. photon eta 2.500000
 max R_gam-clust 0.150000
 min R_isol 0.400000
 R for energy deposit 0.200000
 max E_dep for isolation 10.000000
 smearing on

************************************
*                                  *
*      ***********************     *
*      ***    analyse::Jet   ***   *
*      ***********************     *
*                                  *
************************************

 clusters definition ....
 R cone 0.400000
 jets definition ....
 E_T_jets [GeV] 10.000000
 eta coverage jets 5.000000
 smearing on
 B-field on
```

```
 ************************************
 *                                  *
 *      ************************    *
 *      ***   analyse::Mis   ***    *
 *      ************************    *
 *                                  *
 ************************************

 muon coverage
 min. muon p_T 6.000000
 max. muon eta 2.500000
 unused cells ...
 smearing on
 cells threshold 10.000000
 invisible particles ...
 KF code for invis 66

 ************************************
 *                                  *
 *      ************************    *
 *      ***   analyse::BJet  ***    *
 *      ************************    *
 *                                  *
 ************************************

 ... jets labeling ...
 labeling on/off on
 bjets ...
 min b-quark p_T 5.000000
 max b-quark eta 2.500000
 max R_bj for b-jets 0.200000

 ************************************
 *                                  *
 *      ************************    *
 *      ***   analyse::CJet  ***    *
 *      ************************    *
 *                                  *
 ************************************

 ... jets labeling ...
 labeling on/off on
 cjets ...
 min c-quark p_T 5.000000
 max c-quark eta 2.500000
 max R_cj for c-jets 0.200000

 ************************************
 *                                  *
 *      ************************    *
 *      ***   analyse::Tau   ***    *
 *      ************************    *
 *                                  *
 ************************************

 ... tau-jets labeling ...
 labeling on/off on
 tau-jets ...
 min tau-had p_T 10.000000
 max tau-had eta 2.500000
 max R_tauj for tau-jets 0.300000
 tau-had frac. of jet 0.900000
```

```
*******************************************
*                                         *
*     *******************************     *
*     ***   analyse::Calibration  ***     *
*     *******************************     *
 jets calibration ....
 calibration on
```

## G.3   Control histograms: XXX.conf.root file

Below is the list of control histograms which monitors performance of the simulation/reconstruction. The are store together with `ACDTree` tuple in file XXX.conf.root.

```
 OBJ: TTree    ACDTree  ACDTree : 0 at: 0x8190c80
 KEY: TH1F     hist10000;1     Cell: multiplicity
 KEY: TH1F     hist10101;1     Cluster: multiplicity
 KEY: TH1F     hist10111;1     Cluster: delta eta cluster barycentre
 KEY: TH1F     hist10112;1     Cluster: delta phi cluster barycentre
 KEY: TH1F     hist10113;1     Cluster: delta r   cluster barycentre
 KEY: TH1F     hist10123;1     Cluster: delta r   cluster HPparton
 KEY: TH1F     hist10114;1     Cluster: pTclu/SumpT particle
 KEY: TH1F     hist10124;1     Cluster: pTclu/SumpT HP parton
 KEY: TH1F     hist10210;1     Muon: muon multiplicity NOISOLATED
 KEY: TH1F     hist10211;1     Muon: muon multiplicity ISOLATED
 KEY: TH1F     hist10221;1     Muon: muon multiplicity HP
 KEY: TH1F     hist10231;1     Muon: muon multiplicity HP+ISOL
 KEY: TH1F     hist10311;1     Electron: multiplicity ISOLATED
 KEY: TH1F     hist10321;1     Electron: multiplicity HP
 KEY: TH1F     hist10331;1     Electron: multiplicity HP+ISOL
 KEY: TH1F     hist10411;1     Photon: photon multiplicity ISOLATED
 KEY: TH1F     hist10421;1     Photon: photon multiplicity HP
 KEY: TH1F     hist10431;1     Photon: photon multiplicity HP+ISOL
 KEY: TH1F     hist10501;1     Jet: multiplicity
 KEY: TH1F     hist10511;1     Jet: delta phi jet-barycentre
 KEY: TH1F     hist10512;1     Jet: delta eta jet-barycentre
 KEY: TH1F     hist10513;1     Jet: delta r jet-barycentre
 KEY: TH1F     hist10514;1     Jet: delta r jet- HP parton
 KEY: TH1F     hist10523;1     Jet: pTjet/pT particles in Rcone
 KEY: TH1F     hist10524;1     Jet: pTjet/pT HP parton
 KEY: TH1F     hist10611;1     Mis: reconstructed p_T
 KEY: TH1F     hist10612;1     Mis: reconstructed p_T + cells
 KEY: TH1F     hist10613;1     Mis: reconstructed pTmiss
 KEY: TH1F     hist10621;1     Mis: true p_T invisible
 KEY: TH1F     hist10622;1     Mis: pT miss (true - reco)/reco
 KEY: TH1F     hist10711;1     BJet: b-jets multiplicity
 KEY: TH1F     hist10721;1     BJet: b-quarks HP multiplicity
 KEY: TH1F     hist10723;1     BJet: delta r bjet-bquark HP
 KEY: TH1F     hist10724;1     BJet: pTbjet/pTbquark HP
 KEY: TH1F     hist10811;1     CJet: c-jets multiplicity
 KEY: TH1F     hist10821;1     CJet: c-quarks HP multiplicity
 KEY: TH1F     hist10823;1     CJet: delta r cjet-cquark HP
 KEY: TH1F     hist10824;1     CJet: pTcjet/pTcquark HP
 KEY: TH1F     hist10911;1     Tau: tau-jets multiplicity
 KEY: TH1F     hist10921;1     Tau: tau-had multiplicity
 KEY: TH1F     hist10923;1     Tau: delta r tau-jet, tau-had
 KEY: TH1F     hist10924;1     Tau: pTtaujet/pTtau-had
 KEY: TH1F     hist11001;1     Calibration: calibration correction
 KEY: TH1F     hist11011;1     Calibration: pT jets before calibration
 KEY: TH1F     hist11012;1     Calibration: pT jets after calibration
 KEY: TTree    ACDTree;1       ACDTree
```